\documentclass[english,nofootinbib,reprint,superscriptaddress,preprintnumbers]{revtex4}
\usepackage[T1]{fontenc}
\usepackage[latin9]{inputenc}
\usepackage{babel}
\usepackage{float}
\usepackage{amsmath}
\usepackage{amssymb}
\usepackage{graphicx}
\usepackage{esint}
\usepackage[unicode=true,
 bookmarks=false,
 breaklinks=false,pdfborder={0 0 1},backref=section,colorlinks=false]
 {hyperref}

\makeatletter
\@ifundefined{textcolor}{}
{%
 \definecolor{BLACK}{gray}{0}
 \definecolor{WHITE}{gray}{1}
 \definecolor{RED}{rgb}{1,0,0}
 \definecolor{GREEN}{rgb}{0,1,0}
 \definecolor{BLUE}{rgb}{0,0,1}
 \definecolor{CYAN}{cmyk}{1,0,0,0}
 \definecolor{MAGENTA}{cmyk}{0,1,0,0}
 \definecolor{YELLOW}{cmyk}{0,0,1,0}
}

\usepackage{bbm}
\usepackage{hyperref}
\usepackage{cleveref}

\makeatother

\begin{document}

\preprint{NORDITA-2015-44}

\title{Higuchi ghosts and gradient instabilities in bimetric gravity}

\author{Frank Könnig}

\email{koennig@thphys.uni-heidelberg.de}

\affiliation{Institut für Theoretische Physik, Ruprecht-Karls-Universität Heidelberg,
Philosophenweg 16, 69120 Heidelberg, Germany}

\affiliation{Nordita, KTH Royal Institute of Technology and Stockholm University,
Roslagstullsbacken 23, 10691 Stockholm, Sweden}
\begin{abstract}
Bimetric gravity theories allow for many different types of cosmological
solutions, but not all of them are theoretically allowed. In this
work we discuss the conditions to satisfy the Higuchi bound and to
avoid gradient instabilities in the scalar sector at the linear level.
We find that in expanding universes the ratio of the scale factors
of the reference and observable metric has to increase at all times.
This automatically implies a ghost-free helicity-2 and helicity-0
sector and enforces a phantom dark energy. Furthermore, the condition
for the absence of gradient instabilities in the scalar sector will
be analyzed. Finally, we discuss whether cosmological solutions can
exist, including exotic evolutions like bouncing cosmologies, in which
both the Higuchi ghost and scalar instabilities are absent at all
times.
\end{abstract}
\maketitle

\section{Introduction}

The question whether the graviton can have a mass has been asked for
a long time and its answer has always been accompanied by uncertainties.
The linear theory of a massive gravity was first analyzed by Fierz
and Pauli \cite{FierzPauli1939}. Since then the van Dam-Veltman-Zakharov
discontinuity \cite{vanDamVeltman1970,Zakharov1970} and the appearance
of the Boulware-Deser (BD) ghost \cite{BoulwareDeser1972} has been
challenging the theory. Recently, a theory of a massive spin-2 field
was presented in which the coupling between an additional fixed tensor
field and the metric has a specific structure and is free of the BD
ghost \cite{deRhamGabadadzeTolley2011,deRhamGabadadze2010,deRhamGabadadzeTolley2011b,Hassan2012d,HassanRosen2011a,HassanRosen2012b,HassanRosen2012c,HassanRosenSchmidt-May2012}
(see Refs. \cite{Hinterbichler2011,deRham2014} for recent reviews
on massive gravity). To promote this theory of a massive gravity to
a bimetric theory, Hassan and Rosen considered a dynamical tensor
field $f_{\mu\nu}$ where its kinetic term has the same Einstein-Hilbert
structure as $g_{\mu\nu}$ and does not introduce the BD ghost \cite{HassanRosen2012,HassanRosen2012c}.
This bimetric theory is described by the action
\begin{eqnarray}
S & = & -\dfrac{1}{2}\int d^{4}x\sqrt{-g}R(g)-\dfrac{1}{2}\int d^{4}x\sqrt{-f}R(f)+\int d^{4}x\sqrt{-g}\sum_{n=0}^{4}\beta_{n}e_{n}(X)+\int d^{4}x\sqrt{-g}\mathcal{L}_{m},\label{eq:action_bigravity}
\end{eqnarray}
where we already set the Planck mass for $f_{\mu\nu}$ to $M_{g}$
(see Refs. \cite{Berg2012,HassanSchmidt-MayStrauss2012} for further
explanation), absorbed $m$, the mass scale of the graviton, into
$\beta_{n}$ and expressed masses in units of $M_{g}^{2}$. The interaction
between both tensor fields is determined by the elementary symmetric
polynomials $e_{n}$ of the eigenvalues of the matrices $X_{\gamma}^{\alpha}\equiv\sqrt{g^{\alpha\beta}f_{\beta\gamma}}$,
multiplied by arbitrary real coupling constants $\beta_{n}$. It is
convenient to express these free parameters in units of the present
Hubble expansion rate, $H_{0}^{2}$.

A remarkable property of bimetric gravity theories is the possibility
of nonstandard, self-accelerating cosmological solutions and the ability
of making predictions that are different from $\Lambda$CDM. Some
of these might be useful for future measurements in order to distinguish
standard $\Lambda$CDM from bigravity. To benefit from that, one has
to pay the price and needs to disentangle all the nonviable models
from the viable ones.

Even though this theory has five free parameters, it is not clear
whether viable models exist (except for $\beta_{1}=...=\beta_{4}=0$
which is simply $\Lambda$CDM) and, if they do, what they look like.
In Ref. \cite{Koennig2013} simple criteria of viability were considered
and viable background solutions were presented (see also Ref. \cite{Hassan2014}).
One choice of the coupling parameters, in the following simple \emph{model},
will usually lead to several different cosmological solutions \cite{Koennig2013,Strauss2012,Volkov2012,Berg2012,Comelli2012a,Hassan2014}.
In the following, every possible solution will be called a \emph{branch}.
We distinguish between different types of branches, depending on how
the ratio of the scale factors $r$ of the metrics $f_{\mu\nu}$ and
$g_{\mu\nu}$ evolves. In solutions on \emph{finite branches} the
ratio evolves from zero towards a finite asymptotic value, whereas
on \emph{infinite branches} $r$ becomes infinitely large at early
times and decreases with time. We call all other branches \emph{exotic
branches}, these usually describe bouncing cosmologies or a static
universe in the asymptotic past or future.

So far, only finite and infinite branches were studied in the literature.
While many of these are in good agreement with observational data
at the background level \cite{Strauss2012,Akrami2012,Koennig2013},
most of them suffer from scalar instabilities \cite{Koennig2014b,Koennig2014a,Comelli2012b,DeFelice2014}.
It seems that only one specific class of models, the infinite-branch
bigravity (IBB), is free of scalar instabilities \cite{Koennig2014b}.
These models are specific infinite branch solutions in which $\beta_{2}$
and $\beta_{3}$ vanish. Moreover, IBB agrees very well with observations
at the background and linear level \cite{Koennig2014b,Enander2015}.
Unfortunately, the authors in Ref. \cite{Lagos2014} noted that the
Higuchi bound is generally violated in the early time limit. This
bound, first derived in Ref. \cite{Fasiello2013}, ensures a healthy
helicity-0 mode of the graviton. A violation leads to the appearance
of the Higuchi ghost, named after Higuchi who found that a spin-2
particle with mass $m$ and $0<m^{2}<2H^{2}$ in a de Sitter space
leads to a negative norm \cite{Higuchi1986,Higuchi1989} (see also
Ref. \cite{Fasiello2012} in which the Higuchi bound was derived for
arbitrary spatially flat FLRW metrics in massive gravity). Note that
even though IBB seems to be well behaved at the linear level, the
appearance of the Higuchi ghost may only be visible at higher orders
or maybe even only in the full solution \cite{Woodard2006}. Furthermore,
it was found that cosmological solutions on this infinite branch suffer
from a ghost in the helicity-2 sector at early times \cite{Cusin2014}.

\begin{figure}[H]
\centering{}\includegraphics[width=0.4\textwidth]{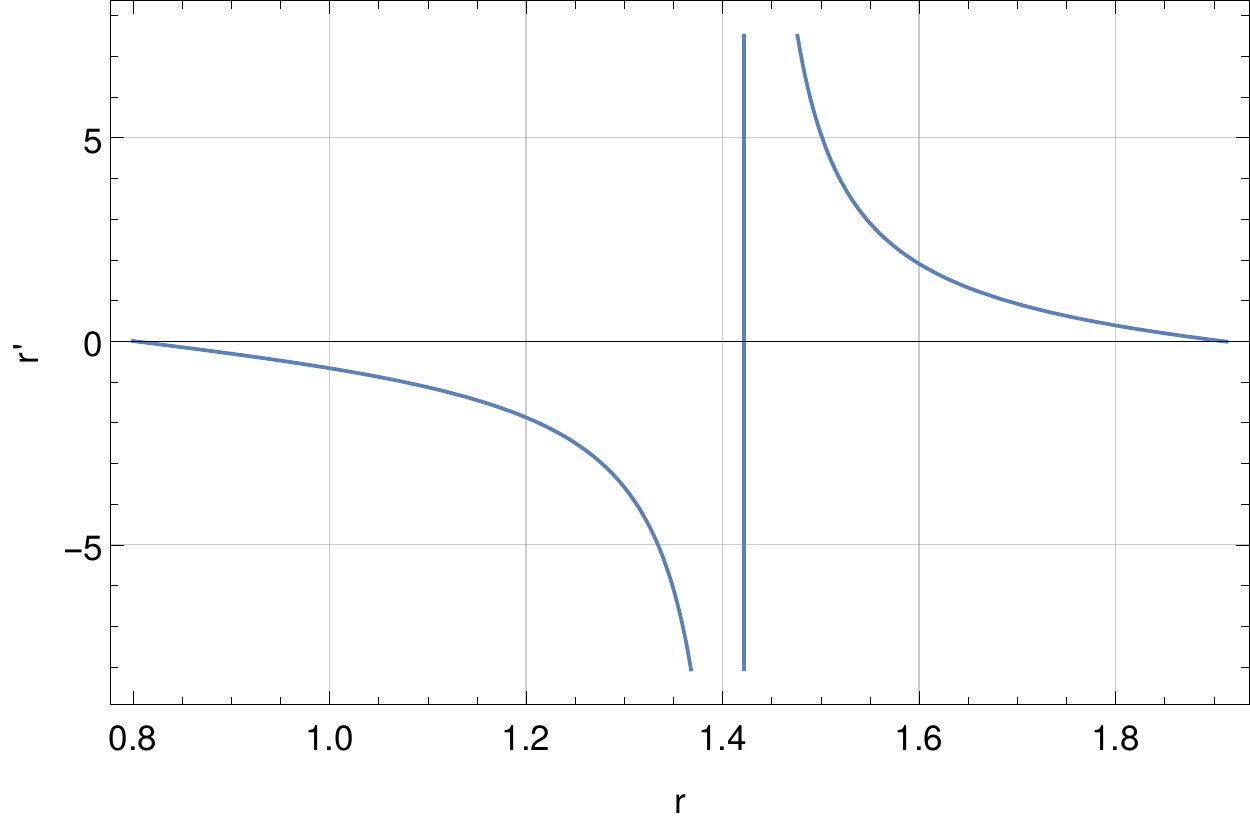}$\qquad\qquad$\includegraphics[width=0.4\textwidth]{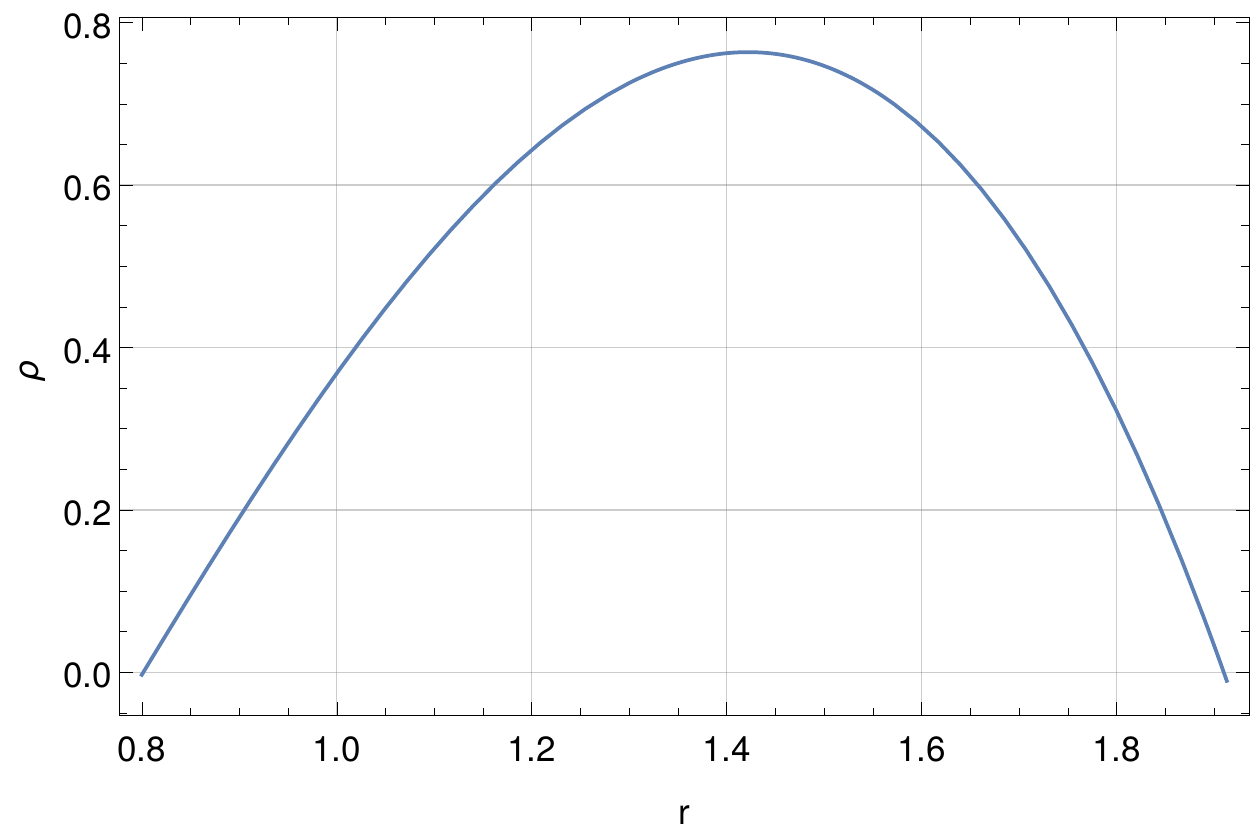}\label{fig:bouncing_model}\protect\caption{Example of a model ($\beta_{i}=\left(0,\,0.3,\,-0.8,\,1,\,-1\right)$)
that describes a bouncing universe. Here, the asymptotic past of this
universe is described by a root at $r\simeq0.8$. It then contracts,
i.e. $dt<0$, until $r$ reaches the pole and, finally, expands towards
a root at $r\simeq1.9$, which describes a de Sitter point.}
\end{figure}

The analysis of viable backgrounds in Ref. \cite{Koennig2013}, that
leaded e.g. to the exclusion of solutions on the exotic branch or
a vanishing $\beta_{2}$ and $\beta_{3}$ in the infinite branch,
are, however, based on assumptions like the existence of a matter
dominated past or the absence of poles in $r'$, where the prime indicates
the derivative with respect to the $e$-folding time $t$. Even though
it would be probably difficult to get exotic solutions in agreement
with observations, they are a priori not excluded. Moreover, poles
in $r'=\frac{d}{dt}r$ could have a very physical meaning: If $r'$
reaches a pole, then $dt$ becomes zero and the Universe undergoes
a bounce. Such an example model is shown in Fig. \ref{fig:bouncing_model}
\footnote{Note that this specific model is not viable due to a negative $\mathcal{H}^{2}$
and is only shown for motivation purposes.%
}.

In the following, we will first briefly discuss the background evolution
in Sec. \ref{sec:Equation-of-Motion}, before we then analyze conditions
for the absence of the Higuchi ghost and scalar instabilities (Sec.
\ref{sec:Higuchi-Bound}-\ref{sec:Eigenfrequencies-of-Scalar}) to
draw conclusions about the viability of all theoretically possible
solutions (Sec. \ref{sec:Finding-Viable-Branches}). 

\section{Equations of motion at background level\label{sec:Equation-of-Motion}}

To find the cosmological background evolution, we vary the action
(\ref{eq:action_bigravity}) with respect to both metrics and find
the equations of motion,
\begin{align}
R_{\mu\nu}-\dfrac{1}{2}g_{\mu\nu}R+\dfrac{1}{2}\sum_{n=0}^{3}\left(-1\right){}^{n}\beta_{n}\left[g_{\mu\lambda}Y_{(n)\nu}^{\lambda}\left(\sqrt{g^{\alpha\beta}f_{\beta\gamma}}\right)+g_{\nu\lambda}Y_{(n)\mu}^{\lambda}\left(\sqrt{g^{\alpha\beta}f_{\beta\gamma}}\right)\right]= & T_{\mu\nu},\\
\bar{R}_{\mu\nu}-\dfrac{1}{2}f_{\mu\nu}\bar{R}+\dfrac{1}{2}\sum_{n=0}^{3}\left(-1\right){}^{n}\beta_{4-n}\left[f_{\mu\lambda}Y_{(n)\nu}^{\lambda}\left(\sqrt{f^{\alpha\beta}g_{\beta\gamma}}\right)+f_{\nu\lambda}Y_{(n)\mu}^{\lambda}\left(\sqrt{f^{\alpha\beta}g_{\beta\gamma}}\right)\right] & =0,
\end{align}
where the overbar denotes curvature of $f_{\mu\nu}$ and $Y_{(n)\nu}^{\lambda}$
are suitable polynomials (see Ref. \cite{Strauss2012} for their definitions).
At the background level, we will use a Friedmann-Lemaître-Robertson-Walker
(FLRW) ansatz for both metrics with two different scale factors, $a$
and $b$, together with two different time parametrizations $t$ and
$\tilde{t}\equiv Xt$. Throughout this work, $t$ represents the $e$-folding
time and a prime denotes the derivative to it. With this ansatz for
the metrics, 
\begin{align}
g_{\mu\nu}dx^{\mu}dx^{\nu} & =a^{2}\left(-\mathcal{H}^{-2}dt^{2}+d\vec{x}^{2}\right),\\
f_{\mu\nu}dx^{\mu}dx^{\nu} & =b^{2}\left(-X^{2}\mathcal{H}^{-2}dt^{2}+d\vec{x}^{2}\right),
\end{align}
where $\mathcal{H}$ is the dimensionless conformal Hubble function,
we obtain the $g_{00}$ and $f_{00}$ equations
\begin{align}
3\mathcal{H}^{2} & =a^{2}\left(\rho+\beta_{0}+3\beta_{1}r+3\beta_{2}r^{2}+\beta_{3}r^{3}\right),\\
3\mathcal{H}^{2} & =\frac{a^{2}rX^{2}}{\left(r'+r\right)^{2}}\left(\beta_{1}+3\beta_{2}r+3\beta_{3}r^{2}+\beta_{4}r^{3}\right).
\end{align}
Here we introduced the ratio of the scale factors $r\equiv b/a$.
As usual, both the Friedmann and acceleration equations for $g_{\mu\nu}$
are degenerated with the conservation of the energy,
\begin{equation}
\rho'=-3\rho\left(1+w_{tot}\right),\label{eq:bigravity_energy_conservation}
\end{equation}
where $w_{tot}$ denotes the equation of state (EOS) parameter of
the total energy density, while there is no extra constraint from
the acceleration equation for $f_{\mu\nu}$ due to the missing coupling
to the energy-momentum tensor. The combination of this set of equations
leads to
\begin{equation}
X=1+\frac{r'}{r}.
\end{equation}
Replacing this constraint in the equations of motion yields
\begin{align}
3\mathcal{H}^{2} & =a^{2}\left(\rho+\beta_{0}+3\beta_{1}r+3\beta_{2}r^{2}+\beta_{3}r^{3}\right),\label{eq:Hubble_bigravity_g}\\
3\mathcal{H}^{2} & =\frac{a^{2}}{r}\left(\beta_{1}+3\beta_{2}r+3\beta_{3}r^{2}+\beta_{4}r^{3}\right).\label{eq:Hubble_bigravity_f}
\end{align}
The second alternative Friedmann equation is particularly interesting
since it directly determines the evolution of the scale factor if
the evolution of $r$ is known.

The sign of $b$ is \emph{a priori} unknown and, therefore, $r$ could
be negative. However, odd powers of $r$ are always proportional to
either $\beta_{1}$ or $\beta_{3}$. All cosmological solutions with
negative $r$ due to a negative scale factor for $f_{\mu\nu}$ are
therefore equivalent to those with positive $r$ after the redefinition
$\beta_{2n+1}\rightarrow-\beta_{2n+1}$. From now on, we will assume
$r\geq0$ %
\footnote{This assumption might only be unjustified if both positive and negative
values of $r$ are reached at some time. In the later discussion we
will find that this requires finite, nonzero values of $r'$ at $r=0$
in order to produce viable branches. It turns out that these specific
models will not be able to produce viable cosmologies.%
}. The combination of both Friedmann equations leads to an equation
for the density as a function of $r$ only,
\begin{equation}
\rho=\beta_{1}r^{-1}-\beta_{0}+3\beta_{2}+3\left(\beta_{3}-\beta_{1}\right)r+\left(\beta_{4}-3\beta_{2}\right)r^{2}-\beta_{3}r^{3}.\label{eq:bigravity_rho}
\end{equation}
It will be useful to study $r'$, which can be written as
\begin{equation}
r'=\frac{\rho'}{\rho_{,r}}=-3\left(1+w_{tot}\right)\frac{\rho}{\rho_{,r}},\label{eq:r_prime}
\end{equation}
where we used Eq. (\ref{eq:bigravity_energy_conservation}) in the
last step.

\section{Higuchi ghosts\label{sec:Higuchi-Bound}}

Bimetric theories are called ghost-free since the specific structure
of the potential term in the Lagrangian avoids an additional degree
of freedom (d.o.f.), which usually would be the BD ghost. This, however,
does not imply that all d.o.f. of the massless and massive graviton
are not ghosts.

\subsection{Higuchi bound}

Bimetric gravity theories describe a mixture of a massless and massive
spin-2 field. The latter carries five dofs, including one helicity-0
mode. In pure massive gravity around a de Sitter spacetime, Higuchi
derived a bound for the graviton mass to ensure positive norm states
\cite{Higuchi1986,Higuchi1989}. A negative norm would imply a ghost
helicity-0 mode and is usually dubbed an Higuchi ghost. The condition
for its absence in bimetric gravity theories around a FLRW background
was derived in Ref. \cite{Fasiello2013}.%
\footnote{Note that the authors in Ref. \cite{Fasiello2013} used an overall
factor of $\frac{1}{2}$ in front of the potential term in the Lagrangian,
which can be compensated for by a redefinition of the $\beta$-couplings.%
} In our notations, the bound is 
\begin{equation}
\frac{3}{2}\left(\beta_{1}+2\beta_{2}r+\beta_{3}r^{2}\right)\left(1+r^{2}\right)\geq\beta_{1}+3\beta_{2}r+3\beta_{3}r^{2}+\beta_{4}r^{3}=3r\left(\frac{\mathcal{H}}{a}\right)^{2},\label{eq:Higuchi_bound_bigravity_in_terms_of_r1}
\end{equation}
 which is equivalent to
\begin{equation}
\beta_{1}+3r^{2}\left(\beta_{1}-\beta_{3}\right)+2r^{3}\left(3\beta_{2}-\beta_{4}\right)+3r^{4}\beta_{3}\geq0.\label{eq:Higuchi_bound_bigravity_in_terms_of_r2}
\end{equation}
Interestingly, using Eqs. (\ref{eq:bigravity_rho})-(\ref{eq:r_prime})
leads to the simple bound
\begin{equation}
\rho_{,r}\leq0.\label{eq:Higuchi_bound_bigravity_in_terms_of_drho/dr}
\end{equation}
This condition for the absence of the Higuchi ghost was already derived
in Ref. \cite{Yamashita2014b} (see also Ref. \cite{DeFelice2014}).
Since
\begin{equation}
\rho_{,r}=-3\left(1+w_{tot}\right)\frac{\rho}{r'}\label{eq:conservation_energy}
\end{equation}
and $\rho>0$ together with $1+w_{tot}>0$ (we are usually considering
a combination of pressureless and relativistic matter), the bound
is equivalent to 
\begin{equation}
r'\geq0.\label{eq:Higuchi_bound_bigravity_in_terms_of_r'}
\end{equation}
Note that this holds even for negative values of $r$. Therefore,
in an expanding universe the ratio of the scale factors $b$ and $a$
has to increase at all times in order to satisfy the Higuchi bound.
Since $r'$ is negative on all infinite branches \cite{Koennig2013},
this directly shows that these branches suffer from the Higuchi ghost
at all times and confirms the findings in Ref. \cite{Lagos2014} that
the bound is violated at least at early times on infinite branches,
i.e. large $r$. On the other hand, all finite branches that produce
viable backgrounds are free from the Higuchi ghost since viability
in these branches enforces $r'\geq0$ \cite{Koennig2013}. This especially
includes the finite branch in the $\beta_{1}$ model, i.e. only $\beta_{1}\neq0$,
which was already shown to be free of the ghost in Ref. \cite{Fasiello2013}.

The rhs of the bound (\ref{eq:Higuchi_bound_bigravity_in_terms_of_r1})
has to be non-negative at all times. Since we already concluded that
$r\geq0$ is a valid assumption without loss of generality, the Higuchi
bound enforces
\begin{equation}
B_{2}\equiv\beta_{1}+2\beta_{2}r+\beta_{3}r^{2}\geq0,\label{eq:def_B2}
\end{equation}
where $B_{2}$ is simply the derivative of $\rho_{mg}$, the modified
part in the Friedmann equation (\ref{eq:Hubble_bigravity_g}), with
respect to $r$. Therefore, the Higuchi bound is related to the change
of the amount of dark energy in our Universe with time.

\subsection{Phantom dark energy}

It is often useful to study the equation of state parameter (EOS),
$w_{mg}$, i.e. the ratio between the pressure and the density, from
contributions of the modification of gravity. If we know how the matter
density in our Universe evolves, then the knowledge of $w_{mg}$ enables
us to draw conclusions about the acceleration and even the future
of our Universe.

In Ref. \cite{Koennig2013} we showed that Eq. (\ref{eq:def_B2})
is directly related to the EOS via
\begin{equation}
w_{mg}=-1-\frac{B_{2}}{\rho_{mg}}r'.
\end{equation}
If $\rho_{mg}>0$ (which, as observations indicate, should hold at
least around present time), then the Higuchi bound enforces a phantom
dark energy. Every cosmological solution in bimetric gravity should
therefore have either a Higuchi ghost or a phantom dark energy.

The property of being a phantom is usually thought to come along with
a future instability, the ``big rip'' \cite{Caldwell2003}. Note,
however, that the EOS is highly time dependent and tends to $-1$
in the asymptotic future if it described by a root in $r'$, e.g.
in most of the finite branch models. A sufficiently fast increase
of $w_{mg}$ could then avoid this instability and guarantee a better
behaved future. A phantom in bimetric gravity is, therefore, not as
frightening as in $\text{\ensuremath{\Lambda}CDM}$. Thus, a model
implying a phantom dark energy should not automatically be related
with a problematic future, much less be rejected.

\subsection{Tensor ghosts}

Interestingly, the only factor in the lapse of $f_{\mu\nu}$ that
is not strictly positive is $r+r'$. Thus, the only way to get a negative
lapse is a negative $r'$. Therefore, fulfilling the Higuchi bound
implies a nonvanishing and especially positive lapse at all times.

It was mentioned in Ref. \cite{Cusin2014} that the relative factor
between the kinetic tensor modes for $g_{\mu\nu}$ and $f_{\mu\nu}$
is the lapse function of $f_{\mu\nu}$ and, therefore, a negative
lapse is responsible for a ghost in the helicity-2 sector. We conclude
that the absence of the Higuchi ghost automatically implies the absence
of a ghost in the helicity-2 sector.

As shown in Ref. \cite{Amendola2015}, the lapse of $f_{\mu\nu}$
directly enters in the friction and mass term of the $f_{\mu\text{\ensuremath{\nu}}}$-tensor
perturbation equation leading to negative values at early times, which
is responsible for a fast grow of the tensor modes \cite{Cusin2014,Amendola2015}.
This is already a signal of the existence of a ghost. To get such
a fast growth in the tensor evolution in accordance with observations
is a challenging but not undoable task \cite{Amendola2015}. The main
problem, however, is the existence of the ghost itself.

\subsection{Consequences of the existence of ghosts}

A ghost helicity-0 or helicity-2 will have a dramatic impact on the
viability of a theory. It will lead to an unbounded Hamiltonian from
below and allow the existence of particles with positive and negative
energies. As expounded in Ref. \cite{Woodard2006}, the vacuum state
will immediately decay into positive and negative energy particles.
This behavior is enough to rule out the underlying theory. %
\footnote{Note that there are ``good ghosts'', e.g. the Faddeev-Popov ghost,
which are not related to physical degrees of freedom and are, therefore,
harmless.%
} It is, therefore, not a question of how problematic the evolution
is of a field described by the equation of motion. A ghost might influence
its evolution in a (more or less) unacceptable way, e.g. through a
negative friction. However, it is not the possibly ill-behaved solution
of the perturbation equations that renders the theory unphysical,
but rather the absence of a stable vacuum state and interactions with
negative energy particles. It is even possible that such a system
could seem to be completely well behaved at all orders in perturbation
theory, but the perturbative solution still not converge to the exact
solution. An example where perturbation theory is even able to hide
the negative energy solutions, which are present in the full theory,
is discussed in Ref. \cite{Woodard2006}.

Since bimetric gravity is only an effective field theory, one might
wonder whether a ghost could be harmless in this setup or whether
a ghost is necessarily excited. This is, unfortunately, not the case.
As explained in Ref. \cite{Sbisa2014}, only modes with positive energy
are able to decouple, but not a ghost state since there is no positive
energy necessary to excite a ghost (see also Ref. \cite{Hassan2014b}).
Even in effective field theories (and even if the mass of the ghost
lies above the cutoff) one has to avoid ghosts at all costs.

\section{Eigenfrequencies of scalar perturbations\label{sec:Eigenfrequencies-of-Scalar}}

After reducing the number of possible cosmological solutions with
the demand of the absence of ghosts, we will analyze the behavior
of scalar perturbations at the linear level. Even though there are
already quite a number of works in which similar properties were studied,
all these investigations were based on strong assumptions, mostly
a restriction in the parameter space, fixing the EOS of the matter
fluid, or focusing on a specific type of branch. In the majority of
cases, this is a consequence of the complexity of the perturbation
equations. Since the aim of this work is to draw conclusions about
the viability of the most general cosmological solutions in bimetric
theories with a FLRW background, we will now work out conditions for
the absence of gradient instabilities without resigning from generality
regarding the parameter space, type of branch and nature, i.e., EOS,
of the fluid.

The set of scalar perturbation equations at the linear level can be
reduced to a system of two second-order differential equations for
two potentials $\Xi_{i}$ describing the two propagating scalar degrees
of freedom \cite{Comelli2012b} (see also Refs. \cite{Koennig2014a,Berg2012,Solomon2014b,Lagos2014,Cusin2014,Comelli2012a}),
\begin{equation}
\Xi_{i}''+A_{ij}\Xi_{j}'+B_{ij}\Xi_{j}=0,
\end{equation}
where $A_{ij}$ and $B_{ij}$ are matrices which depend on the background
quantities $r$, $\mathcal{H}$ and the parameters of the models.
The complexity of this system depends crucially on the choice of the
gauge. A very convenient one was used in Ref. \cite{Lagos2014}, leaned
on Ref. \cite{Lagos2013}. In this work, we take advantage of the
relatively simple%
\footnote{Where ``simple'' means that printing these equations would fill
only a couple of pages. %
} perturbation equations that the authors found in this gauge (see
Ref. \cite{Lagos2014} for the derivation and printed equations) and
analyze them by using the ansatz $\Xi_{i}\propto e^{\omega t}$. For
simplicity, we assume that the eigenfrequencies $\omega$ do not depend
on time. This is a valid assumption as long as $|\omega'/\omega^{2}|\ll1$
holds and was confirmed for all models studied in Ref. \cite{Koennig2014b}.
In the subhorizon limit, we obtain a surprisingly simple expression
for the eigenfrequencies, 
\begin{align}
\omega^{2} & =\left(\frac{k}{\mathcal{H}}\right)^{2}\left[\frac{r'\left(\frac{\left(r^{2}+1\right)\left(\beta_{1}-\beta_{3}r^{2}\right)r'}{\rho(w+1)}-\frac{r^{2}\left(\beta_{1}+4\beta_{2}r+3\beta_{3}r^{2}\right)}{\beta_{1}+2\beta_{2}r+\beta_{3}r^{2}}\right)}{3r^{3}}-1\right]\\
 & =\left(\frac{k}{\mathcal{H}}\right)^{2}\left[\frac{r'\left(-\rho_{,r}^{-1}\left(r^{2}+1\right)\left(\beta_{1}-\beta_{3}r^{2}\right)-\frac{r^{2}\left(\beta_{1}+4\beta_{2}r+3\beta_{3}r^{2}\right)}{\beta_{1}+2\beta_{2}r+\beta_{3}r^{2}}\right)}{3r^{3}}-1\right],\label{eq:eigenfreq}
\end{align}
which agrees with all previous, but much more complicated, results
for one- and two-parameter models that were studied in \cite{Koennig2014b}.
As already mentioned in Ref \cite{Koennig2014b}, if we assume dark
matter only, then for models in which $\beta_{2}=\beta_{3}=0$ this
reduces to 
\begin{equation}
\omega_{\beta_{0}\beta_{1}\beta_{4}}^{2}=\left(\frac{k}{\mathcal{H}}\right)^{2}\frac{r''}{3r'}.\label{eq:eigenfrequencies_b0b1b4}
\end{equation}
In order to discuss stability, we only need to analyze the sign of
$\omega^{2}$: A negative value would imply oscillating and, therefore,
stable potentials $\Xi_{i}$. If, however, $\omega^{2}$ is positive,
then $\Xi_{i}$ grows quickly with time and even faster as the scales
become smaller. Such an instability is not compatible with the structure
in our Universe and needs to be avoided in a viable model.

Let us now introduce $B_{2}=\beta_{1}+2\beta_{2}r+\beta_{3}r^{2}$
to obtain

\begin{align}
\omega^{2} & =\frac{k^{2}}{3r\rho_{,r}\mathcal{H}^{2}}\left[r'\left(3\left(r^{2}+1\right)\left(\frac{B_{2}}{r}\right)_{,r}-\rho_{\text{,r}}\left(r\frac{B_{2,r}}{B_{2}}+1\right)\right)-3r\rho_{,r}\right].
\end{align}
Interestingly, the condition for stability depends on how dark energy
(and the density of the cosmic fluid) changes but not explicitly on
how large it is. We observed a similar property during the analysis
of the Higuchi bound. Note that $B_{2}$ is related to the change
of the energy density, $\rho_{,r}$, and the Hubble expansion via\emph{
\begin{equation}
B_{2}=-\frac{r}{1+r^{2}}\left(\frac{1}{3}r\rho_{,r}-2\left(\frac{\mathcal{H}}{a}\right)^{2}\right).\label{eq:relation_between_B2_and_drhodr}
\end{equation}
}Together with
\begin{align}
\left(\frac{B_{2}}{r}\right)_{,r}= & r^{-2}B_{2}\left(r\frac{B_{2,r}}{B_{2}}-1\right),
\end{align}
we finally arrive at
\begin{align}
\omega^{2} & =\left(\frac{k}{\mathcal{H}}\right)^{2}\left(\frac{2r'\left(r\left(r^{2}+1\right)B_{2}{}_{\text{,r}}\rho_{\text{,r}}-\left(\frac{\mathcal{H}}{a}\right)^{2}\left(3\left(r^{2}+1\right)B_{2}{}_{\text{,r}}+r\rho_{\text{,r}}\right)+6\left(\frac{\mathcal{H}}{a}\right)^{4}\right)}{r^{2}\rho_{\text{,r}}\left(r\rho_{\text{,r}}-6\left(\frac{\mathcal{H}}{a}\right)^{2}\right)}-1\right).\label{eq:eigenfreq_in_terms_of_B2,r_rho,r}
\end{align}
As we will see later, this expression for the eigenfrequencies will
become very convenient when analyzing the stability around poles in
$r'$, which e.g. always appear in exotic branches.

It might be useful to study an expression for $\omega^{2}$ which
does not explicitly depend on the $\beta$ parameters but on $r$
and its derivatives, like Eq. (\ref{eq:eigenfrequencies_b0b1b4}).
Finding such an expression is always possible when using a set of
five independent equations to eliminate all coupling parameters. One
possibility is the set of equations for $r'$, $r''$, $r'''$, $\mathcal{H}^{2}$
and $\rho$ (note that the result will not depend on $r'''$) which
yields
\begin{equation}
\omega^{2}=\left(\frac{k}{\mathcal{H}}\right)^{2}\frac{a^{2}\rho r^{2}(w+1)\left[2(w+1)r''+r'\left(6w^{2}-2w'+9w+3\right)\right]-2\mathcal{H}^{2}r'\left[r'\left((w+1)\left(r'-3rw\right)+rw'\right)-r(w+1)r''\right]}{3r(w+1)r'\left(a^{2}\rho r(w+1)+2\mathcal{H}^{2}r'\right)}.
\end{equation}
Here, and in all the following equations, we dropped the subscript
in $w_{tot}$ for simplicity. If we are interested in analyzing the
eigenfrequencies at specific epochs, e.g. radiation dominated era
(RDE) and matter dominated era, we can assume $w\simeq\text{const}$
and obtain
\begin{equation}
\omega^{2}=\left(\frac{k}{\mathcal{H}}\right)^{2}\frac{a^{2}\rho r^{2}(w+1)\left[2r''+3r'(2w+1)\right]+2\mathcal{H}^{2}r'\left[r\left(r''+3wr'\right)-r'^{2}\right]}{3rr'\left(a^{2}\rho r(w+1)+2\mathcal{H}^{2}r'\right)}.
\end{equation}
This leads to the condition
\begin{align}
r'\left[a^{2}\rho r(w+1)+2\mathcal{H}^{2}r'\right]\left[a^{2}\rho r^{2}(w+1)\left(2r''+(6w+3)r'\right)+2\mathcal{H}^{2}r'\left(r\left(r''+3wr'\right)-r'^{2}\right)\right] & <0\label{eq:condition_scalar_stability}
\end{align}
in order to get stable scalar perturbations, i.e. $\omega^{2}<0$.
When using the Higuchi bound, $r'>0$, the first bracket term is always
positive and, thus, the second one has to be negative. This is equivalent
to
\begin{equation}
r''<\frac{r'}{2r}\frac{2\mathcal{H}^{2}r'\left(r'-3rw\right)-3a^{2}\rho r^{2}(w+1)(2w+1)}{a^{2}\rho r(w+1)+\mathcal{H}^{2}r'},\label{eq:condition_scalar_stability_r''}
\end{equation}
where we also used $r'>0$. Note that the denominator is always positive.
If the numerator would be negative, then the bound would especially
imply $r''<0$. However, this is not generally the case and, thus,
the condition for stable scalar modes is not automatically equivalent
to $r''<0$ in contrast to the case for $\beta_{0}\beta_{1}\beta_{4}$
models during matter domination (see Eq. (\ref{eq:eigenfrequencies_b0b1b4})).

\subsection{Radiation-dominated era}

Even though we will not aim to exclude models which are theoretically
allowed but do very likely not reproduce observational data (an example
would be a nearly static universe that did not have a radiation-dominated
epoch), it is worthwhile to analyze the conditions when the Universe
is filled with either relativistic particles or pressureless matter
only.

When radiation dominates, i.e. $w\simeq1/3$, the eigenfrequencies
simplify to
\begin{align}
\omega^{2}= & \frac{k^{2}}{3\mathcal{H}^{2}rr'}\frac{\frac{4}{3}a^{2}\rho r^{2}\left(2r''+5r'\right)+2\mathcal{H}^{2}r'\left(rr''-r'^{2}+rr'\right)}{\frac{4}{3}a^{2}\rho r+2\mathcal{H}^{2}r'}.
\end{align}
In the early Universe, the Hubble expansion is usually driven by radiation,
i.e. $3\mathcal{H}^{2}\simeq a^{2}\rho$. With this approximation,
the condition for stability in the scalar sector becomes 
\begin{align}
r'' & >-\frac{r'\left(r'+10r\right)}{r'+4r}.\label{eq:condition_scalar_stability_RDE}
\end{align}
For large absolute values of $r'$, which is the case e.g. near a
pole, we simply obtain $r''>-r'$.

In a previous work \cite{Koennig2014b}, we studied the eigenfrequencies
for IBB and confined ourselves to a universe filled with dark matter
only. According to Eq. (\ref{eq:eigenfrequencies_b0b1b4}), we concluded
stable scalar modes because $r'$ increases with time but stays negative
until reaching the final de Sitter point. Since $r'$ is always negative
in IBB, the condition (\ref{eq:condition_scalar_stability_RDE}) is
not necessarily valid anymore. However, we can still use condition
(\ref{eq:condition_scalar_stability}). Here, the product of the first
two terms is always positive since
\begin{equation}
r'\left(a^{2}\rho r(w+1)+2\mathcal{H}^{2}r'\right)\Big|_{\text{IBB}}=9a^{2}\beta_{1}r\left(r^{2}+1\right)\left(\frac{(w+1)\left(\beta_{1}+\beta_{4}r^{3}-3\beta_{1}r^{2}\right)}{\beta_{1}-2\beta_{4}r^{3}+3\beta_{1}r^{2}}\right)^{2}>0.
\end{equation}
Therefore, we can analyze the third factor and, assuming $w\in\left(-1,1\right)$
for simplicity, find that stable modes are guaranteed if
\begin{equation}
3\beta_{1}r^{2}<\beta_{1}+\beta_{4}r^{3},\label{eq:condition_stability_IBB}
\end{equation}
which is not only satisfied in the RDE, i.e. large $r$ (note that
both $\beta_{1}$ and $\beta_{4}$ have to be positive in order to
get a viable cosmological background), but, in fact, is equivalent
to the condition $\rho>0$ on that branch and, therefore, trivially
satisfied at all times.

\subsection{Matter-dominated era}

Let us study the regime when matter dominates the Universe. Now the
EOS vanishes and the scalar modes are described through
\begin{align}
\omega^{2}= & \frac{k^{2}}{3\mathcal{H}^{2}rr'}\frac{a^{2}\rho r^{2}\left(2r''+3r'\right)+2\mathcal{H}^{2}r'\left(rr''-r'^{2}\right)}{a^{2}\rho r+2\mathcal{H}^{2}r'}.
\end{align}
For stability, we need to satisfy the condition
\begin{align}
r'' & <\frac{2\mathcal{H}^{2}r'^{3}-3a^{2}\rho r^{2}r'}{2a^{2}\rho r^{2}+2r\mathcal{H}^{2}r'}.
\end{align}
 If we assume that $\frac{a^{2}\rho}{\mathcal{H}^{2}}\rightarrow0$
for late times, which should be true when dark energy starts to dominate,
then the condition of stability reduces to
\begin{equation}
r''\lesssim\frac{r'^{2}}{r}.
\end{equation}

\section{Finding Viable Branches\label{sec:Finding-Viable-Branches}}

We will now raise the question whether branches exist that satisfy
both the Higuchi bound and the condition for scalar stability. Here
we will only focus on cosmological solutions that are not equivalent
to $\Lambda$CDM, which of course satisfy both conditions. We therefore
assume that at least one of the couplings $\beta_{1},...,\beta_{4}$
is nonzero. Together with conditions of physicality, $a,\rho$, $\mathcal{H}^{2}>0$,
we define these as criteria of viability. Note that we allow for solutions
that have a very nonstandard past, e.g. no matter- or radiation-dominated
epoch, or even contracting backgrounds, even though these might be
hard to compare with observational data. This extends the more restrictive
background analysis of \cite{Koennig2013}. Therefore, not only the
finite branch with small $r$ or the infinite one could be viable
but also many solutions on exotic branches. Many different types of
branches exist: some of them start from a root $r'=0$, while others
may evolve from a pole or even pass a pole at some finite time. In
many cases it is not directly clear whether such branches solve the
equations of motion. In particular, every branch always needs to contain
a solution of Eq. (\ref{eq:Hubble_bigravity_f}) at present time,
i.e. when $\mathcal{H}=a=1$.

We start with focusing on finite branches with a root at $r=0$. Let
us first concentrate on models with $\beta_{1}\neq0$, which always
have a root at $r=0$ (see Eq. (\ref{eq:r_prime})). In Ref. \cite{Koennig2014b}
we generally found scalar instabilities in these type of branches.
Even though this is based on the assumption of a universe filled with
dark matter only, this conclusion does not change when considering
arbitrary but reasonable EOS parameters. We take the same line of
argument and study the simple $\beta_{1}$-models, i.e. models with
only nonvanishing $\beta_{1}$, since all other models will reduce
to these in the limit when $r$ gets close to $r=0$. The eigenfrequencies
in $\beta_{1}$ models are given by

\begin{equation}
\omega_{\beta_{1}}^{2}=\frac{1+2w-6r^{2}(w+2)-9r^{4}}{\left(3r^{2}+1\right)^{2}}\left(\frac{k}{\mathcal{H}}\right)^{2}\simeq\frac{1+2w}{\left(3r^{2}+1\right)^{2}}\left(\frac{k}{\mathcal{H}}\right)^{2}
\end{equation}
and, therefore, indicate unstable modes for small values of $r$ as
long $w>-1/2$. Let us consider the previously excluded models with
$\beta_{1}=0$ and find 
\begin{equation}
r'\Big|_{r=0}=\frac{\beta_{0}-3\beta_{2}}{\beta_{3}}\left(w+1\right).
\end{equation}
Even though the combination $\beta_{0}=3\beta_{2}$ is able to produce
a root at $r=0$, it will not lead to viable solutions since in this
case $r'=-3\left(1+w\right)+\mathcal{O}\left(r^{2}\right)$ indicates
a violation of the Higuchi bound. From this we conclude that
\begin{enumerate}
\item[\emph{1.}] \emph{Finite branches with a root at $r=0$ always lead to either
unstable modes (if $\beta_{1}\neq0$) or violate the Higuchi bound
(if $\beta_{1}=0$) for small $r$.}
\end{enumerate}
On the other hand, $r'$ could be nonzero but still finite at $r=0$.
In this case, one of the asymptotic points is either a pole or the
whole branch evolves between two roots at negative and positive $r$.
In the first case, we can assume that at least one of the poles is
reached at $r>0$, otherwise we are able to analyze viability in the
``mirrored'' model corresponding to $\beta_{2n+1}\rightarrow-\beta_{2n+1}$.
If the branch does not contain any pole, then $\rho_{,r}$ has to
vanish at $r=0$ (roots at $r\neq0$ always indicate a vanishing density
whereas a maximum of the density at $r\neq0$ leads to a pole). The
position of the maximum of $\rho$ at $r=0$ requires $\beta_{3}=0$
and leads to $\rho_{,r}\propto r$ which cannot be negative for both
regions, $r>0$ and $r<0$. We can summarize that
\begin{enumerate}
\item[\emph{2.}] \emph{All finite branches with a nonzero and finite $r'$ at $r=0$
have to have a pole either in the asymptotic past or future.}
\end{enumerate}
Roots, except for those at $r=0$, always correspond to a vanishing
density. Due to Eq. (\ref{eq:r_prime}), we will always find a pole
between two roots $r_{1}$ and $r_{2}$, if both $r_{1}$ and $r_{2}$
are nonzero. Therefore, poles could be interesting starting or final
points of a branch. Whenever such a pole describes the asymptotic
future, then $\dot{r}$ has to go to zero, otherwise the pole would
not be a stable asymptotic point. Since $r'=\mathcal{H}^{-1}\frac{d}{d\tau}r=\mathcal{H}^{-1}a\dot{r}$
diverges, we find that $\mathcal{H}$ needs to vanish at this point.
On the other hand, if a pole describes the asymptotic past, then we
can use the fact that the density starts from a finite value. For
nonzero values of $r$, this is clear from Eq. (\ref{eq:bigravity_rho}).
It also holds if $r=0$ is a pole, since this would require $\beta_{1}=\beta_{3}=0$
due to Eq. (\ref{eq:r_prime}) \cite{Koennig2013} and, therefore,
implies $\rho\big|_{r=0}=3\beta_{2}-\beta_{0}$. If the density is
finite at early times, the scale factor $a$ has to have a finite
but nonzero value. In this case, $\mathcal{H}$ needs to be zero at
early times, too, otherwise one could go backwards in time and we
would not have an asymptotic past. Thus, we conclude
\begin{enumerate}
\item[\emph{3.}] \emph{$\mathcal{H}$ has to become zero on a pole, if it describes
an asymptotic point.\label{Lemma_H_vanishes_on_pole}}
\end{enumerate}
Let us assume a pole at $r=0$, which, as we already noted, requires
$\beta_{1}=\beta_{3}=0$ and leads to $\mathcal{H}^{2}\big|_{r=0}=\beta_{2}a^{2}$.
From the previous conclusion, we need a vanishing $\mathcal{H}^{2}$
at $r=0$. Note that $a>0$, otherwise this would contradict a finite
density. Therefore, we need $\beta_{2}=0$ and, thus, obtain $B_{2}=0$
for all $r$, which means that
\begin{enumerate}
\item[\emph{4.}] \emph{A pole at $r=0$ violates the Higuchi bound.}
\end{enumerate}
For simplicity, we will from now on assume that if there is a pole
at $r_{p}$, then $r_{p}>0$. Additionally, we can exclude $r=0$
from being an asymptotic point due to the previous conclusions.\textbf{
}Furthermore, Eqs. (\ref{eq:r_prime}) and (\ref{eq:bigravity_rho})
provide the limit $r'\propto-r$ when taking $r\rightarrow\infty$
as long as the density does not vanish (see Ref \cite{Koennig2013}
more detailed explanations). This excludes infinite branches, i.e.
branches in which $r$ evolves from or to $r\rightarrow\infty$, from
being viable due to the violation of the Higuchi bound and we find
that
\begin{enumerate}
\item[\emph{5.}] \emph{The limits $r\rightarrow0$ and $r\rightarrow\infty$ are no
viable asymptotic points.}
\end{enumerate}
We will now consider a root at $r\neq0$ as the asymptotic past. Due
to Eq. (\ref{eq:r_prime}), the density vanishes on a root. To fulfill
the conservation of energy, those models require a contracting universe
at early times. If this universe evolves to another root (on which
again $\rho=0$), then it has to undergo a bounce at $\rho_{,r}=0$
leading to a pole at which $\mathcal{H}=0$. Employing the previous
conclusions, we find the general statement
\begin{enumerate}
\item[\emph{6.}] \emph{Every viable branch needs to contain at least one pole on which
$\mathcal{H}$ vanishes.}
\end{enumerate}
This result is particularly interesting as it will allow us to draw
conclusions when connecting this with the requirement of stable scalar
perturbations and the absence of the Higuchi ghost. The necessary
condition for a pole is $\rho_{,r}\rightarrow0$. Then, the eigenfrequencies
of scalar perturbations around the pole (\ref{eq:eigenfreq_in_terms_of_B2,r_rho,r})
reduce to
\begin{align}
\omega^{2} & \rightarrow\left(\frac{k}{\mathcal{H}}\right)^{2}\left(2B_{2}{}_{\text{,r}}\frac{r'\left(1+r^{2}\right)}{r^{2}\rho_{\text{,r}}}\frac{r\rho_{\text{,r}}-3\left(\frac{\mathcal{H}}{a}\right)^{2}}{r\rho_{\text{,r}}-6\left(\frac{\mathcal{H}}{a}\right)^{2}}-1\right)\\
 & \simeq\left(\frac{k}{\mathcal{H}}\right)^{2}\left(2B_{2}{}_{\text{,r}}\frac{r'}{\rho_{\text{,r}}}\frac{\left(1+r^{2}\right)}{r^{2}}-1\right)\label{eq:eigenfrequencies_pole}
\end{align}
where we used $\left(\frac{\mathcal{H}}{a}\right)^{4}\ll\left(\frac{\mathcal{H}}{a}\right)^{2}$,
$\left(\frac{\mathcal{H}}{a}\right)^{2}\rho_{,r}\ll\left(\frac{\mathcal{H}}{a}\right)^{2}$
(and, additionally, $B_{2,r}\neq0$ which, as we will see later, is
justified), together with 
\begin{equation}
\frac{r\rho_{\text{,r}}-3\left(\frac{\mathcal{H}}{a}\right)^{2}}{r\rho_{\text{,r}}-6\left(\frac{\mathcal{H}}{a}\right)^{2}}=\frac{B_{2}\left(1+r^{2}\right)-\left(\frac{\mathcal{H}}{a}\right)^{2}r}{B_{2}\left(1+r^{2}\right)}\simeq1,
\end{equation}
which follows from Eq. (\ref{eq:relation_between_B2_and_drhodr})
and $B_{2}\left(1+r^{2}\right)-\left(\frac{\mathcal{H}}{a}\right)^{2}r\simeq B_{2}\left(1+r^{2}\right)$
(note that the Higuchi bound (\ref{eq:Higuchi_bound_bigravity_in_terms_of_r1})
implies $B_{2}\left(1+r^{2}\right)>2r\left(\frac{\mathcal{H}}{a}\right)^{2}>0$).
Since $r'\rightarrow\infty$ and $\rho_{,r}\rightarrow0$ (but still
$\rho_{,r}<0$ and $r'>0$), the first term in the bracket of Eq.
(\ref{eq:eigenfrequencies_pole}) dominates unless $B_{2,r}=0$. 

Let us first assume that $B_{2,r}=0$ at the pole $r_{p}$. We then
find
\begin{align}
B_{2,r}\Big|_{r=r_{p}}=0 & \quad\Rightarrow\quad\beta_{2}=-\beta_{3}r_{p},\\
\left(\frac{\mathcal{H}}{a}\right)^{2}\Big|_{r=r_{p}}=0 & \quad\Rightarrow\quad\beta_{1}=-\beta_{4}r_{p}^{3},\\
\rho_{,r}\Big|_{r=r_{p}}=0 & \quad\Rightarrow\quad\beta_{3}=-\beta_{4}r_{p},
\end{align}
which leads to
\begin{equation}
3\mathcal{H}^{2}=\frac{a^{2}}{r}\beta_{4}\left(r-r_{p}\right)^{3},
\end{equation}
as well as
\begin{equation}
B_{2}=-\beta_{4}r_{p}(r_{p}-r)^{2}.
\end{equation}

If $r$ increases with time (which implies that $dt>0$ since $r'>0$),
then $\beta_{4}$ has to be positive in order to get a positive $\mathcal{H}^{2}$.
On the other hand, this would imply a negative $B_{2}$ and, thus,
would violate the Higuchi bound. We therefore have to have a decrease
of $r$ with time (which implies contraction, $dt<0$). Now, this
is only compatible with negative values for $\beta_{4}$. Of course,
such a model would be hard to believe in since it would contract at
all times. But there is a more solid argument for ruling out these
models: The contraction would lead to an increasing density. Since
a root corresponds to a vanishing density, there must be a point of
maximum density which always indicates a pole (see Eq. (\ref{eq:r_prime})).
Note, that we already excluded both $r=0$ and $r\rightarrow\infty$
as asymptotic states, which are the only ones that would be able to
describe an infinitely large density. However, on this one, $\mathcal{H}$
cannot vanish. Even though this second pole does not necessarily need
to be an asymptotic point, $\mathcal{H}=0$ is required due to the
bounce. Neither the positive nor the negative values for $\beta_{4}$
lead to viable solutions and we conclude that 
\begin{enumerate}
\item[\emph{7.}] \emph{Viability enforces a nonzero value for $B_{2,r}$ around a
pole.}
\end{enumerate}
We are now allowed to assume $B_{2,r}\neq0$. Then the term $-1$
in Eq. (\ref{eq:eigenfrequencies_pole}) is negligible and, thus,
$B_{2,r}$ has to be positive in order to get stability, i.e. $\omega^{2}<0$. 

We will now study the expansion rate around the pole $r_{p}$ and
check whether $\mathcal{H}^{2}$ is positive. Note that $\left(\mathcal{H}^{2}\right)_{,r}\big|_{r_{p}}$
does not automatically vanish since $\mathcal{H}^{2}$ could become
negative, too (which, however, would not correspond to physical solutions).
However, the conditions for scalar stability ($B_{2,r}\big|_{r_{p}}>0$),
the existence of a pole ($\rho_{,r}\big|_{r_{p}}=0$ with $\mathcal{H}^{2}\big|_{r_{p}}=0$)
and physicality ($\rho\big|_{r_{p}}\geq0$) together with the assumption
that $\left(\mathcal{H}^{2}\right)_{,r}\big|_{r_{p}}\neq0$ lead to
a contradiction. Therefore, let us assume $\left(\mathcal{H}^{2}\right)_{,r}\big|_{r_{p}}=0$
which, together with $\mathcal{H}^{2}\big|_{r_{p}}=0$, implies 
\begin{align}
\beta_{2} & =-\frac{1}{3}r_{p}^{-1}\left(2\beta_{1}-\beta_{4}r_{p}^{3}\right),\\
\beta_{3} & =\frac{1}{3}r_{p}^{-2}\left(\beta_{1}-2\beta_{4}r_{p}^{3}\right).
\end{align}

If we now assume that $\mathcal{H}^{2}$ is positive and nonzero at
second order, then we need to have
\begin{equation}
3\left(\frac{\mathcal{H}}{a}\right)^{2}=\frac{(r_{p}-r)^{2}}{rr_{p}^{2}}\left(\beta_{1}+\beta_{4}rr_{p}^{2}\right)=r_{p}^{-3}\left(\beta_{1}+\beta_{4}r_{p}^{3}\right)\left(r-r_{p}\right)^{2}+\mathcal{O}\left(\left(r-r_{p}\right)^{3}\right)>0.\label{eq:Hubble_pole}
\end{equation}
However, this would imply that
\begin{equation}
\rho_{,r}=2\frac{1+r_{p}^{2}}{r_{p}^{3}}\left(\beta_{1}+\beta_{4}r_{p}^{3}\right)\left(r-r_{p}\right)+\mathcal{O}\left(\left(r-r_{p}\right)^{2}\right)
\end{equation}
only becomes negative when leaving the pole, if $r$ decreases with
time, i.e. in a contracting universe. We can now use the same argument
that we used before and conclude that we need to reach a second pole
which will either describe an asymptotic point or a bounce. Eq. (\ref{eq:Hubble_pole})
provides the possibility of another point $r_{p_{2}}=-\beta_{1}/\left(\beta_{4}r_{p}^{2}\right)$
at which the expansion stops but this cannot be a pole since then
we would find
\begin{equation}
\rho_{,r}\Big|_{r_{p_{2}}}=-\frac{\left(\beta_{1}+\beta_{4}r_{p}^{3}\right){}^{2}\left(\beta_{1}^{2}+\beta_{4}^{2}r_{p}^{4}\right)}{\beta_{1}\beta_{4}^{2}r_{p}^{6}}\neq0.
\end{equation}
Our last chance are models in which $\mathcal{H}^{2}$ vanishes up
to second order implying that
\begin{align}
\beta_{1} & =-\beta_{4}r_{p}^{3},\\
\beta_{2} & =\beta_{4}r_{p}^{2},\\
\beta_{3} & =-\beta_{4}r_{p}.
\end{align}
These solutions lead to $B_{2,r}=0$, which we already excluded earlier.
Therefore,
\begin{enumerate}
\item[8.] \emph{A negative $B_{2,r}$ around a pole leads to gradient instabilities
whereas a positive value violates either the Higuchi bound or leads
to unstable scalar perturbations.}
\end{enumerate}
In combination with the requirement of a positive $B_{2,r}$ in order
to get stable scalar perturbations this shows that every branch is
plagued by either the Higuchi ghost or scalar gradient instabilities.

\section{Conclusions and Outlook}

We analyzed general models in singly coupled bimetric gravity around
a FLRW background and found that all physical cosmological solutions
that are not equivalent to $\Lambda$CDM have a period in time in
which either linear scalar perturbations undergo a gradient instability
or the Higuchi ghost appears. The condition for the absence of ghosts
is surprisingly equivalent to $r'>0$, which means that the ratio
of the scale factors $b$ and $a$ has to increase as long as the
Universe expands. Moreover, satisfying this bound ensures a positive
lapse of $f_{\mu\nu}$ which is related to the absence of a helicity-2
ghost.

\begin{figure}

\includegraphics[width=1\textwidth]{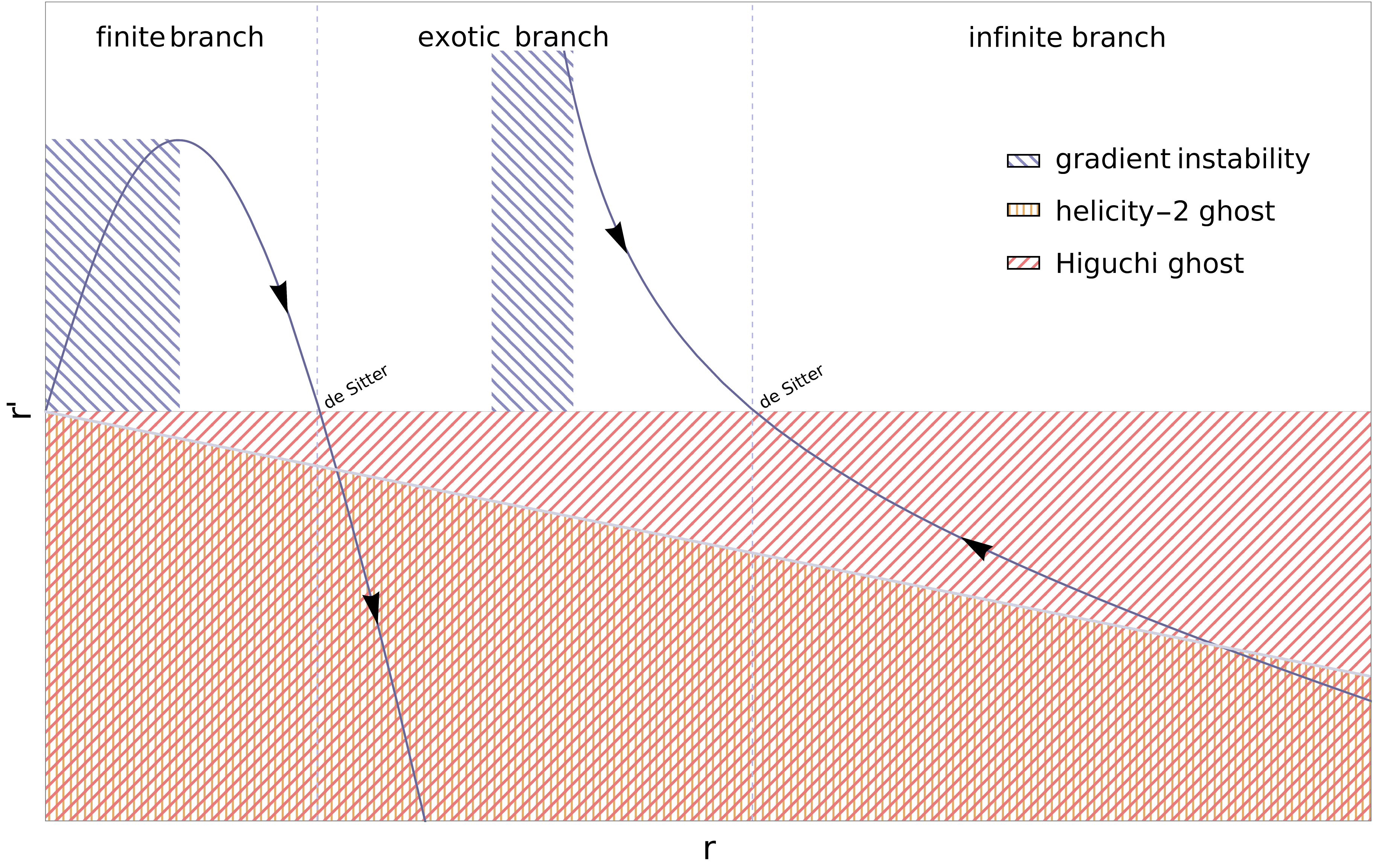}\protect\caption{\label{fig:forbidden_regions}Illustration of a phase space diagram
of a typical model together with colored regions corresponding to
different types of instabilities. While finite branches are plagued
from gradient instabilities in the scalar sector (diagonal blue stripes
from top-left to bottom-right) at early times, the infinite branches
suffer from the Higuchi ghost (diagonal red stripes from bottom-left
to top-right) at all times and a ghost in the helicity-2 sector (vertical
orange stripes) at early times. Finally, all exotic branches, including
e.g. bouncings, have always a pathological behavior at least around
the pole in $r'$.}

\end{figure}

In fact, all infinite branches suffer from the Higuchi ghost at all
times and a ghost in the helicity-2 sector at early times, whereas
in all finite branches, and even exotic branches that do not contain
the limit $r\rightarrow0$, there exists at least one epoch in which
there is either a gradient instability in the scalar sector or a ghost
appears. A schematic illustration of a typical phase space diagram
with the forbidden regions is presented in Fig. \ref{fig:forbidden_regions}.

While the existence of a ghost renders the model unphysical and forces
us to discard this type of model, unstable scalar modes will not necessarily
rule out the theory. A Vainshtein screening may be able to prevent
the scalar sector from getting unstable. Furthermore, this gradient
instability is not present at all times. Every finite branch has a
point in time at which the instability stops and the scalar perturbations
begin to oscillate. As shown in \cite{Akrami2015}, a small, but natural,
Planck mass for $f_{\mu\nu}$ %
\footnote{Note that in this and many previous works, the Planck mass $M_{f}$
was set to $M_{g}$, which is allowed due to a redundancy in the parameters
but is, however, not the most natural choice.%
} would shift this gradient instability to very early times or even
to energy scales above the cutoff of the effective field theory. In
the latter case, the cosmological evolution would be very close to
$\Lambda$CDM. On the other hand, if the instability ended between
inflation and big bang nucleosynthesis, only very small scales would
be affected \cite{Akrami2015}. These could, in principle, lead to
a creation of many seeds for black holes. %
\footnote{Since the cosmological evolution at the time where the instability
would end is not close to $\Lambda$CDM yet, the fast change in $\rho_{mg}$
might even have a stronger influence on the evolution of primordial
black holes compared to standard $\Lambda$CDM \cite{Enander2009}.%
}

All models which we do not have to exclude due to the presence of
a ghost will describe a phantom dark energy. Such a property would
cause an anxious future in a $\Lambda$CDM model but not necessarily
in bimetric theories due the time dependence of EOS corresponding
to dark energy. In fact, it could cause welcome signatures that might
allow observations to distinguish bimetric gravity from general relativity.

Throughout this work we assumed a very simple, but well-motivated,
type of bigravity. We considered a fluid that is only singly coupled
to an observable metric and where both metrics are of FLRW type. Several
extensions exist in the literature. One example would be the coupling
of matter to both metrics $g_{\mu\nu}$ and $f_{\mu\nu}$ simultaneously
\cite{Tamanini2013,Akrami2013b,Akrami2014,Aoki2014,DeFelice2014,Comelli2014},
which, however, would introduce the BD ghost if the same matter sector
is coupled to both metrics \cite{Tamanini2013,deRham2014b,Yamashita2014}.
Ghost-free (but not always with well-behaved cosmological solutions)
scenarios exist if one assumes a coupling through a composite metric
\cite{deRham2014b,deRham2014c,Hassan2014b,Schmidt-May2014,Noller2014,Heisenberg2014,Enander2014,Comelli2015,Gumrukcuoglu:2015nua,Solomon2014}.
But even the bimetric gravity with a standard matter coupling could
allow for cosmological solutions without any gradient or ghost instabilities
at the cost of giving up a FLRW background \cite{Nersisyan2015,Maeda2013}.

\begin{acknowledgments}
I am grateful to Yashar Akrami, Luca Amendola, Jonas Enander, Matteo
Fasiello, Fawad Hassan, Macarena Lagos, Edvard Mörtsell, Angnis Schmidt-May,
and Adam Solomon for fruitful discussions and suggestions. Additionally,
I am thankful to NORDITA (Stockholm) for organizing the ``Extended
Theories of Gravity'' workshop that inspired discussions during the
completion of this work and to both NORDITA and the Oskar Klein Centre
for the warm hospitality during my subsequent visit. I acknowledge
support from the Landesgraduiertenförderung (LGFG) through the Graduate
College ``Astrophysics of Fundamental Probes of Gravity'' and from
the DFG through the Grant No. TRR33, ``The Dark Universe''.
\end{acknowledgments}
\bibliographystyle{apsrev}
\bibliography{HiguchiScalarStability}

\end{document}